\documentclass[runningheads]{llncs}
\usepackage[mobile]{eccv}
\usepackage{eccvabbrv}
\usepackage{graphicx}
\usepackage{booktabs}
\usepackage{multirow}
\usepackage{makecell}
\usepackage{mathabx}
\usepackage[accsupp]{axessibility}  %
\usepackage{wrapfig}
\usepackage[pagebackref,breaklinks,colorlinks,citecolor=eccvblue]{hyperref}
\usepackage{orcidlink}

\definecolor{MyDarkBlue}{rgb}{0,0.08,1}
\definecolor{MyDarkGreen}{rgb}{0.02,0.6,0.02}
\definecolor{MyDarkRed}{rgb}{0.8,0.02,0.02}
\definecolor{MyDarkOrange}{rgb}{0.40,0.2,0.02}
\definecolor{MyPurple}{RGB}{111,0,255}
\definecolor{MyRed}{rgb}{1.0,0.0,0.0}
\definecolor{MyGold}{rgb}{0.75,0.6,0.12}
\definecolor{MyDarkgray}{rgb}{0.66, 0.66, 0.66}
\definecolor{MyDarkCyan}{rgb}{0.05, 0.55, 0.45}
\definecolor{MyBlack}{rgb}{0., 0., 0.}
\definecolor{MyMagenta}{rgb}{1., 0., 1.}
\definecolor{BerkeleyYellow}{RGB}{255,204,41}
\definecolor{BerkeleyLightBlue}{RGB}{94,146,221}
\definecolor{BkDarkBlue}{rgb}{.05,.07,.353}

\newcommand{\ignorethis}[1]{}

\usepackage[title]{appendix}

\begin{document}

\title{supDQN: Supervised Rewarding Strategy Driven Deep Q-Network for sEMG Signal Decontamination} 

\titlerunning{supDQN: Supervised Rewarding Strategy Driven Deep Q-Network for sEMG Signal Decontamination}

\author{Ashutosh Jena\inst{1} \quad Naveen Gehlot\inst{1} \quad Rajesh Kumar\inst{1} \\ Ankit Vijayvargiya\inst{2,3} \quad Mahipal Bukya\inst{4} }

\authorrunning{Ashutosh Jena et al.}

\institute{Malaviya National Institute of Technology\inst{1}\; \;
Dublin City University\inst{2}\; \; Swami Keshvanand Institute of Technology\inst{3}\; \; Manipal Academy of Higher Education\inst{4}}

\maketitle

\begin{abstract}
The presence of muscles throughout the active parts of the body such as the upper and lower limbs, makes electromyography-based human-machine interaction prevalent. However, muscle signals are stochastic and noisy. These noises can be regular and irregular. Irregular noises due to movements or electrical switching require dynamic filtering. Conventionally, filters are stacked, which trims and delays the signal unnecessarily. This study introduces a decontamination technique involving a supervised rewarding strategy to drive a deep Q-network-based agent (supDQN). It applies one of three filters to decontaminate a 1sec long surface electromyography signal, which is dynamically contaminated. A machine learning agent identifies whether the signal after filtering is clean or noisy. Accordingly, a reward is generated. The identification accuracy is enhanced by using a local interpretable model-agnostic explanation. The deep Q-network is guided by this reward to select filter optimally while decontaminating a signal. The proposed filtering strategy is tested on four noise levels (-5 dB, -1 dB, +1 dB, +5 dB). supDQN filters the signal desirably when the signal-to-noise ratio (SNR) is between -5 dB to +1 dB. It filters less desirably at high SNR (+5 dB). A normalized root mean square ($\Omega$) is formulated to depict the difference of filtered signal from ground truth. This is used to compare supDQN and conventional methods including wavelet denoising with debauchies and symlet wavelet, high order low pass filter, notch filter, and high pass filter. The proposed filtering strategy gives an average $\Omega$ value of 1.1974, which is lower than the conventional filters.
\end{abstract}

\section{Introduction}
\label{intro}
Applications of surface electromyography (sEMG) have increased rapidly in the past two decades and still accelerating. According to data gathered from the Web of Science, the number of studies on sEMG per year has gone from 32 in 2003 to 445 in 2022~\cite{WOS}. In these studies, the most commonly observed applications of sEMG signal are rehabilitation, robotic prosthesis, medical diagnosis, gesture-controlled robotics, analysis, medical research, etc~\cite{appl1, appl2, appl3}. While all these studies consider the usage of machine learning and artificial intelligence to achieve the task at hand, the raw sEMG signal requires a highly accurate and precise pre-processing methodology to remove noise. In sEMG signal these noises are usually referred to as noise or contaminants in signal, which can be caused by motion artifact (MOA), physiological interference, powerline interference (PLI), white Gaussian noise (WGN), amplifier saturation, fatigue muscles, etc~\cite{rmoa}. The removal of such contaminants is necessary to avoid deviation from the actual outcome of these studies. Therefore, there is a need to identify and remove these contaminants from an sEMG signal.

Significant efforts have been made to decontaminate an sEMG signal. The most common decontamination techniques used in sEMG signal pre-processing are bandpass filtering (BPF) and notch filtering (NF). BPF is used to reduce the power level of samples having a frequency beyond the range of sEMG signals while NF is used to do the same for a specific frequency which reduces the PLI level in the signal~\cite{moaWgn, pli, crossTalk}. However, these are static filters that act on the entire signal. This can result in a reduction of power levels of the actual sEMG signal. The advanced form of filtering includes wavelet denoising~\cite{43}, decompositional filtering techniques~\cite{47} and signal whitening~\cite{50}. Wavelet denoising is a premature technique that simply removes coefficients beyond a defined threshold. It is very useful in analysis. While the decompositional filtering techniques are very efficient, they are very difficult to tune. The selection of the mother wavelet and degree of decomposition is a very hectic task. The more robust methods ICA and PCA are introduced in~\cite{45,46,51}. These methods can reliably separate noises of different kinds~\cite{commonFilt}, but only at a high power level of noise. As the level of noise reduces, they fail to identify and hence it becomes difficult to separate noise from clean signal. Most recently, M.C. Tosin et. al formulated noise removal as a reinforcement learning problem in~\cite{steps} and dynamically filtered four different types of noise introduced at different intervals of a signal. They considered an actor-critic agent with a fuzzy logic-based rewarding strategy to achieve optimal filtering of noise. The fuzzy logic-based rewarding strategy requires an inference table which can only be determined by an expert. However, in~\cite{steps}, the authors deduced a heatmap-based fuzzy inference table before the learning process. This heatmap table guides the output of the fuzzy system which is then used to generate a reward. This becomes tedious when a change in noise level is introduced.

This study considers a new approach to simplify and diversify the reward generation process while improving the efficiency at each step of reward generation. At the same time, an agent capable of solving this problem efficiently is obtained, tested, and evaluated. The work done in this study is as follows:
\begin{itemize}
    \item A deep Q-network-based learning agent, that learns to identify and decontaminate an sEMG signal dumped with three different noises at different intervals.
    \item A new and efficient reward generation strategy involving machine learning agents (MLA) is proposed for the decontamination of sEMG signal. To improve the efficiency of the MLA, a local interpretable model-agnostic explanation (LIME) tool is used to interpret and finally select the optimal set of features.
\end{itemize}

\begin{figure}[h]
    \includegraphics[width = \textwidth]{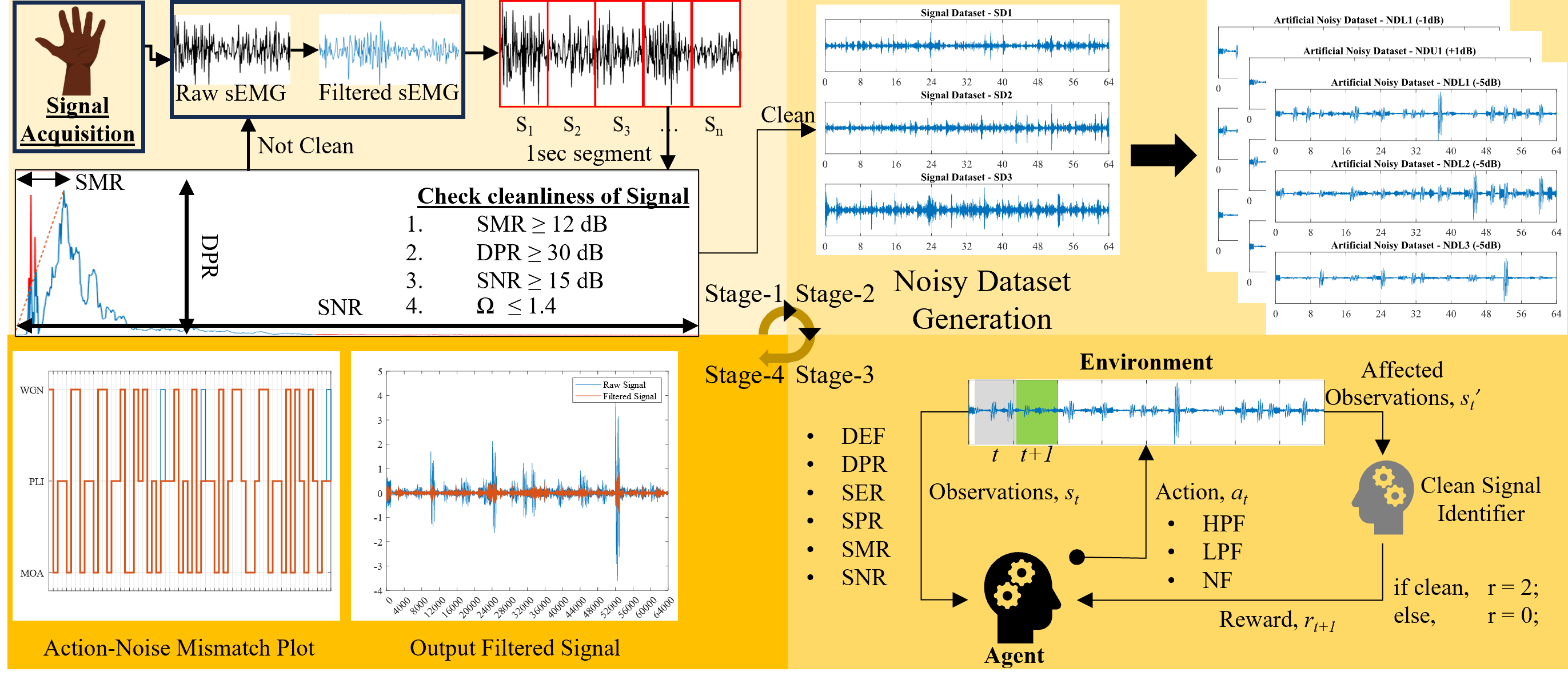}
    \caption{Flow of study: Stages 1 and 2 deal with artificial contamination of a clean sEMG signal. They have been discussed in Section~\ref{sec:data} under the subtopics of signal cleaning and artificial contamination, respectively. Stage-3 deals with the methodology of proposed filtering and it has been discussed in Section~\ref{sec:meth}. Stage 4 is the results and comparative analysis which is discussed in Section~\ref{resAnal}.}
    \label{flowDiag}
\end{figure}

The upcoming sections are organized as follows: The problem is introduced in Section~\ref{intro}. Section~\ref{sec:data} and~\ref{sec:meth} present the foundation of the dataset and the proposed filtering method, respectively. The result with analysis is discussed in Section~\ref{resAnal}. Section~\ref{sec:data} -~\ref{resAnal} are the four stages described in Fig.~\ref{flowDiag} indicating the flow of study. Finally, the study is concluded in Section~\ref{conc}.
\section{Dataset} \label{sec:data}
This section presents the method of artificial generation of noisy sEMG signals for the study. Mathematically, the artificial contamination can be represented as in Eq.~\eqref{artCont}. This equation represents a general idea of the addition of noise ($n(t)$) on top of a clean signal ($x(t)$) at a level decided by the value of $A$ to obtain a noisy sEMG signal ($y(t)$).

\begin{equation} \label{artCont}
    y(t) = x(t) + A \times n(t)
\end{equation}
For the study, three clean signals are obtained by following the procedure described in Section~\ref{sec:sigclean}. Three different types of noise (MOA, PLI, and WGN) are considered which is elaborated in subsection~\ref{sec:noiseAddition}. The other noises are physiological (electrocardiogram (ECG)) and instrumental noise, which are not considered in the study due to a lack of standardization in integrating ECG with sEMG to produce noisy sEMG and a lack of a model to represent instrumental noise. Subsection~\ref{sec:noiseAddition}, also explains the procedure of contamination. The reliability of the filtering agent is checked at the different SNRs: -5 dB, -1 dB, +1dB, and +5 dB. The outcomes of this section are nine noisy sEMG datasets at four different noise levels.
\subsection{Signal Cleaning} \label{sec:sigclean}
\begin{figure}[h]
    \centering
    \includegraphics[width = 0.8\textwidth]{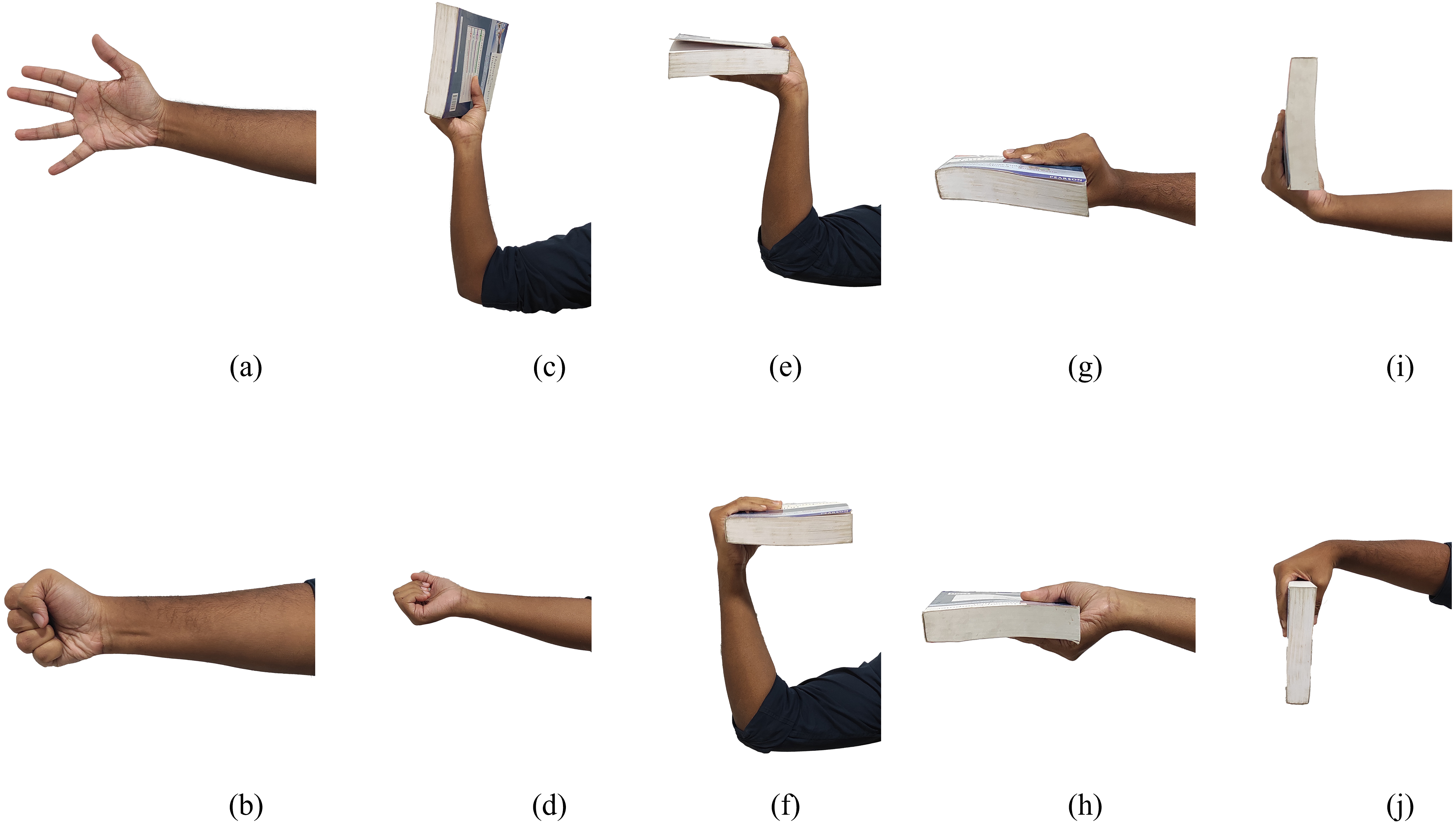}
    \caption{The ten hand activities performed to obtain sEMG signals.}
    \label{fig:handImage}
\end{figure}

sEMG signal is acquired using a BIOPAC MP150 device~\cite{biopac}. The signal is acquired while the subject holds a book for 2 minutes. The acquisition frequency is set at 2 kHz and only one channel is used. For a proper analysis, the above-mentioned procedure is repeated 10 times while holding the book in different ways as shown in Fig.~\ref{fig:handImage}.
Since, the study revolves around the identification and removal of noise, the cleanest possible form of sEMG signal is acquired by setting up the options for low pass filtering. Any value of frequency component, above the range occupied by sEMG signal~\cite{as1} is filtered using 500 Hz LP mode of filtered signal acquisition. The acquired signal, however, is not free of low-frequency MOA and PLI. These noises are filtered using digital filters and a signal purity check is performed using the method described in~\cite{cc}. The purity of the signal is checked over a window of 1sec following the constraints given in Table~\ref{tab:sigCheck}. If the values of SMR, SNR, DEF, and DPR are inside the acceptable range, it is marked `good'. It is intuitively deduced that, if the percentage of total good marked segments over the entire signal is above 90\%, the signal can be assumed suitable. However, if a signal is found impure it is discarded. The pre-processed signal obtained at the end of this stage is referred to as a clean signal in further stages.
\begin{table}[h]
    \centering
    \caption{Constraints to check purity of signal~\cite{cc}}
    \label{tab:sigCheck}
    \begin{tabular}{l|c}
    \hline
        \textbf{Features} & \textbf{Acceptable Range}  \\ \hline \hline
        Signal to motion artifact ratio (SMR) & $SMR \ge 12dB$ \\
        Signal to noise ratio (SNR) & $SNR \ge 15dB$\\
        Power density (DPR) & $DPR \ge 30dB$\\
        Deformation ratio ($DEF$) & $DEF \le 1.4$ \\ \hline
    \end{tabular}
\end{table}
For a generalized analysis, two conditions need to be fulfilled by the clean signal datasets:
\begin{itemize}
    \item The signals to be considered should be different from each other based on activity.
    \item The signals must fulfill the cleanliness criteria as shown in table~\ref{tab:sigCheck}.
\end{itemize}
Initially, 10 raw sEMG datasets are obtained having a length of 96 seconds and after filtering and signal purity check, it is observed that only 1 sEMG dataset (sEMG-10) is suitable in the length of 96 seconds (refer, Table~\ref{tab:sigCheck1}).
\begin{table}[h]
\centering
\caption{Signal cleanliness checked over 96sec long signals}
\label{tab:sigCheck1}

\begin{tabular}{l|ccccc}
\hline
Signals & sEMG-1 & sEMG-2 & sEMG-3 & sEMG-4 & sEMG-5   \\ \hline \hline
Cleanness (\%) & 79.17  & 57.29  & 15.1   & 69.79  & 11.98 \\ \hline
Signals & sEMG-6 & sEMG-7 & sEMG-8 & sEMG-9 & sEMG-10  \\ \hline \hline
Cleanness (\%) & 89.58  & 30.73  & 59.89  & 84.89  & 93.75 \\ \hline
\end{tabular}
\end{table}
Since, considering the 96sec long signal violates the conditions for generalized analysis, two more signals close to 90\% clean are explored (sEMG-6 and sEMG-9). Dividing these three signals (sEMG-6, sEMG-9, and sEMG-10) further into 32sec long signals, it is observed that the conditions are fulfilled (refer, to Table~\ref{tab:sigCheck2}). These three clean signals of length 32sec are selected for the study.
\begin{table}[h]
    \centering
    \caption{Signal cleanliness checked over 32sec intervals}
    \label{tab:sigCheck2}
    \begin{tabular}{l|c|c|c|c}
    \hline
        \textbf{Signals} & \multicolumn{3}{c|}{\textbf{Percent cleanliness (in \%)}} & \textbf{Nomenclature} \\ 
        \cline{2-4}
        & 0-32sec & 32-64sec & 64-96sec & \\ \hline \hline
        sEMG-6 & 87.5 & 85.9375 & \textbf{95.3125} & SD1 \\
        sEMG-9 & 81.25 & \textbf{90.625} & 82.8125 & SD2 \\
        sEMG-10 & \textbf{95.3125} & 93.75 & 92.1875 & SD3\\ \hline
    \end{tabular}
\end{table}
\subsection{Artificial Contamination} \label{sec:noiseAddition}
The three clean sEMG signals are used to obtain nine noisy sEMG signals. The three noises used for doping with a clean signal include MOA, PLI, and WGN. These noises are added randomly to the signal in windows of 500 msec. Three random sequences are prepared, like the one shown in Fig.~\ref{noiseAdd}. Following these three sequences, noises are added to the three clean datasets forming a total of nine noisy sEMG datasets. The information regarding the dataset of noise used during noise addition is described in brief as follows:
\begin{figure}[h]
    \centering
    \includegraphics[width=0.8\textwidth]{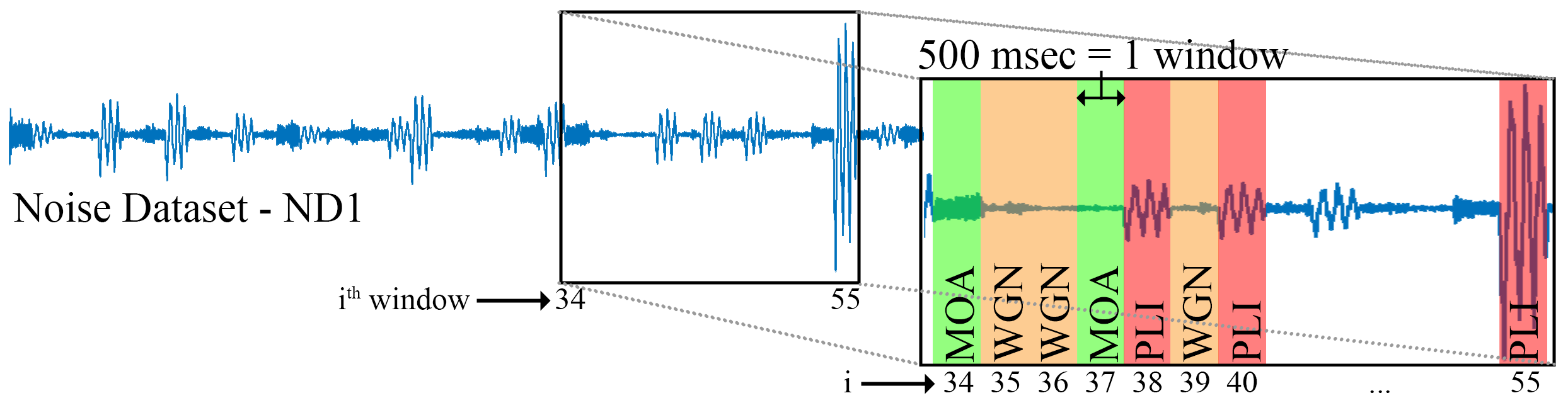}
    \caption{Illustration of noise added in random sequence}
    \label{noiseAdd}
\end{figure}
\subsubsection{Motion Artifact}
MOA is a low-frequency and high-power level artifact with a typical frequency range of 0.5 to 20 Hz. It is a result of skin-electrode interface movement. It cannot be generated, therefore it is collected using the BIOPAC MP150 DAQ setup. However, while collecting the signal for MOA, a series of 50-60 taps are done in the proximity of the electrode-attached muscle in one minute~\cite{steps}. To retain only the signal containing MOA the obtained signal is then filtered beyond 35 Hz using a low pass filter with a cutoff frequency of 35 Hz. The resulting signal is assumed to be a purely MOA noise.
\subsubsection{Powerline Interference}
PLI is an electromagnetic interference occurring due to the electrical devices in the proximity of the acquisition setup. It affects the frequency domain specifically at the powerline frequency. In the study, PLI is a $sine$ function as represented in Eq.~\eqref{e1}
\begin{equation} \label{e1}
    n_{pli}(t) = \sin{(2\pi \times 50t)} + \frac{1}{3}\sin{(2\pi \times 150t)}
\end{equation}
\subsubsection{White Gaussian Noise}
WGN is a Gaussian noise distributed throughout the frequency domain having a low power level. Since it is distributed in the entire frequency range, its filtering is the most difficult. However, ~\cite{rwgn,rmoa} suggests that removal of the signal after 500 Hz should effectively improve the SNR of the signal. Data for WGN is obtained using the $wgn$ function in MATLAB with a random power level between -8 and -4 dB. In Eq.~\eqref{e2}, r is a random variable obtained from the set of real numbers uniformly distributed over the range of -8 and -4 dB.
\begin{subequations}
\begin{align}
    r & \in \mathbb{N} \cap [-8,\: -4] \label{e2}\\
    n_{wgn}(t) & = wgn(r)
\end{align}
\end{subequations}
The final noisy datasets (NDs) are prepared by adding $n_{moa}(t)$, $n_{pli}(t)$ and $n_{wgn}(t)$ to the main signal dataset (SD) in a pre-defined random sequence. This can be mathematically expressed as follows:
\begin{equation}
    y_{ijk}(t) =  x_{j_i}(t) + \alpha (p_{k_i}  n_{moa}(t) + q_{k_i}  n_{pli}(t) + r_{k_i} n_{wgn}(t)) \: \: \forall i
\end{equation}
Here, $p_{k_i}, q_{k_i}, r_{k_i} \in \{0,1\}$ and $p_{k_i} + q_{k_i} + r_{k_i} = 1 \: \forall \: i \in \mathbb{Z}^+ \cap [1, T]$. $x_{j_i}(t)$ represents the 1000 samples long $i^{th}$ segment of the $j^{th}$ clean signal. $T$ represents the total timesteps in an episode. $y_{jk}(t)$ corresponds to the $i^{th}$ segment of final noisy signal obtained out of $j^{th}$ clean signal doped with noise following $k^{th}$ noise sequence. Here, $j, k \in \{1,2,3\}$. `$\alpha$' indicates the level of noise and it is given by the following equation, 
\begin{equation} \label{alp}
    \alpha = \frac{P_c}{P_n \times (10^{0.1 P_{req}}-1)}
\end{equation}
Here, $P_c, \: P_n, \: P_{req}$ represents the power level of the clean signal, noise signal, and required SNR in dB, respectively.
\section{Methodology} \label{sec:meth}
The problem of filtering considered in this study is viewed as a reinforcement learning problem. A DQN agent-based reinforcement learning environment is set up to achieve dynamic filtering of an sEMG signal. Where the goal of the agent is to minimize the noise appearing in the signal by taking suitable filtering action. The subsequent sections deal with the formalization of a reinforcement learning problem out of the noise decontamination problem.
\subsection{sEMG Environment Setup}
The sEMG signal is treated as the environment in which a timestep increment is observed over a window of 500 msec of the signal. At timestep, t = 1 the first 500 msec of the signal becomes the current observational segment and the subsequent 500 msec of the signal becomes the next observational segment for t = 2. Therefore, in a particular episode, a maximum of 64 timesteps are possible. At the end of Section~\ref{sec:noiseAddition}, nine noisy sEMG datasets (NDs) are formed. One out of the nine is used for training (ND1) and the rest are for testing and evaluation (ND2-ND9). After the training, four supDQN filtering agents are formed, each for a specified SNR level.
\subsubsection{Observations}
The physical significance of observation in any environment is the physical values that may represent the situation of the environment at that timestep. In this case, at any timestep, the environment can be represented by deformation ratio (DEF), peak density (DPR), signal-to-motion artifact ratio (SMR), signal-to-noise ratio (SNR), signal-to-powerline ratio (SPR) and signal-to-electrocardiogram ratio (SER). There can be more observations from the environment at any particular timestep, however, their values may not be significant from the perspective of goal. Mathematical formulae for these physical parameters are referred from~\cite{cc, steps}.
\subsubsection{Actions} \label{sec:action}
Actions are the tasks that an agent performs on its environment at any particular time step. In this case, filtering of the raw signal is considered as the action. The different actions that the agent is equipped with are high pass filtering (HPF) against MOA, low pass filtering (LPF) against WGN, and specific frequency notch filtering (NF) against PLI. The filters are digital elliptic filters with an 80 dB attenuation. Elliptic filters are chosen for their better steady-state response even at a reduced order of the system. At any timestep, the agent observes the environment it is currently in and then takes action following a policy. The task of the supDQN agent is to obtain a policy that yields the highest Q-value over an episode. The supDQN agent and the Q-value are defined in Section~\ref{sec:dqn}.
\subsubsection{Rewarding Strategy} \label{sec:rwd}
Reward is the performance measure of an action. To obtain an optimal policy, the reward must be a positive incentive for good and a negative incentive for bad action. It can be represented as a function of state (s) and action (a). However, due to the stochastic nature of an sEMG signal, and its features (i.e. observations at every timestep), it is necessary to use a trained MLA to distinguish good action from bad. Therefore, during the study, the reward for an action is determined by a MLA, trained with the training dataset to identify clean and noisy signals.
\begin{figure}[h]
    \centering
    \includegraphics[width=0.8\textwidth]{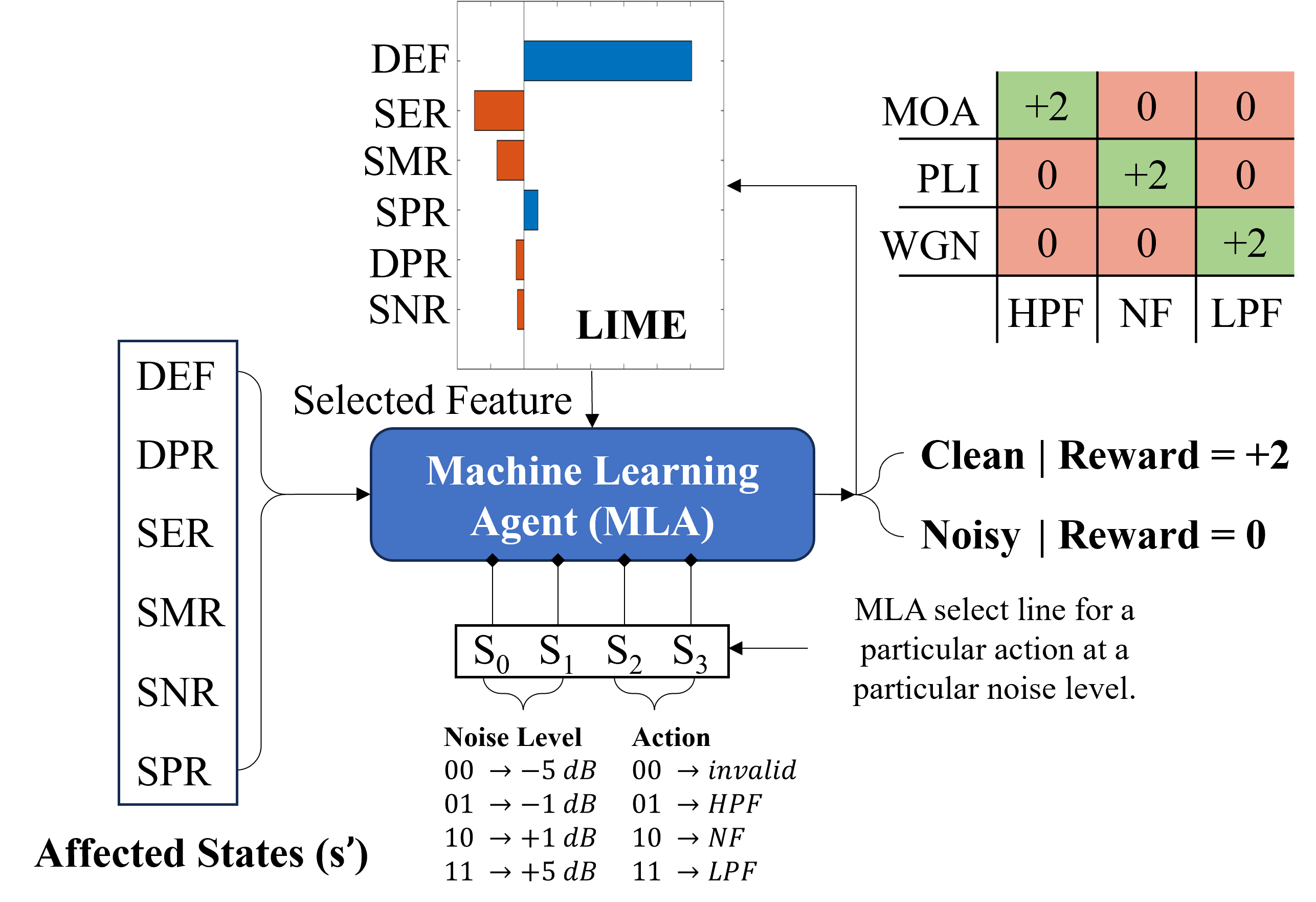}
    \caption{Rewarding Strategy}
    \label{rwds}
\end{figure}
\begin{figure}[t]
    \begin{subfigure}{0.16\linewidth}
        \includegraphics[width=\textwidth]{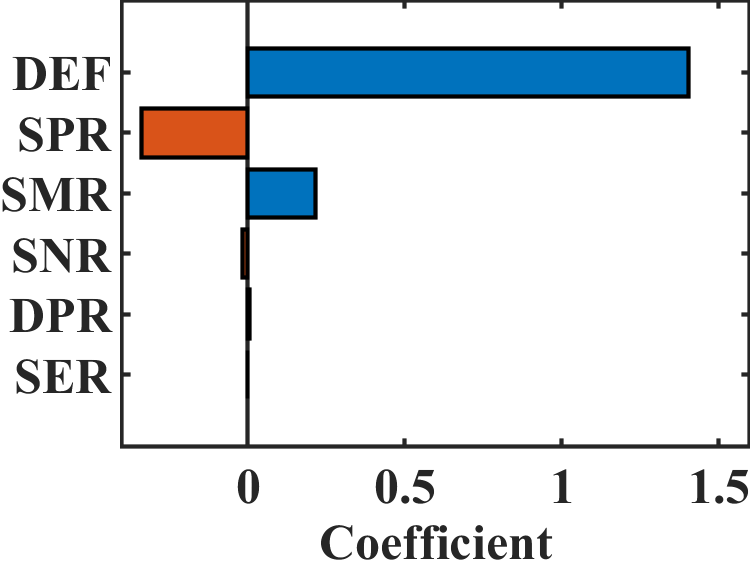}
        \caption{Model-0001}
        \label{xai1}
    \end{subfigure}
    \begin{subfigure}{0.16\linewidth}
        \includegraphics[width=\textwidth]{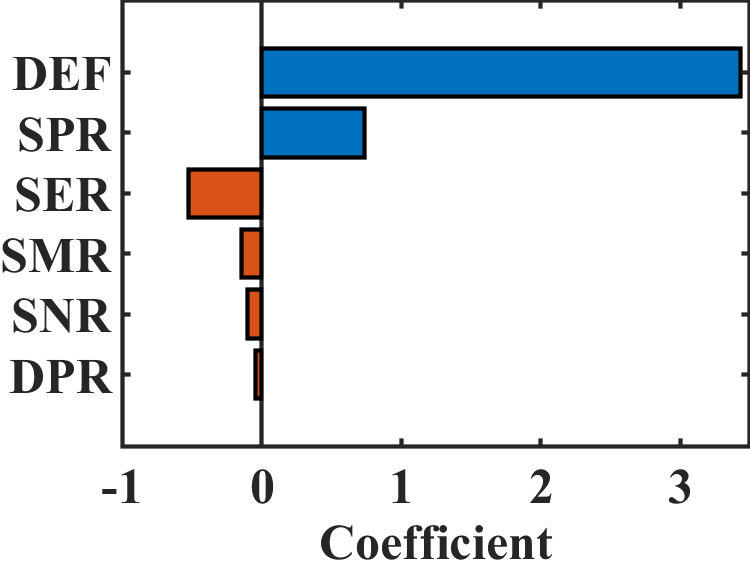}
        \caption{Model-0010}
        \label{xai2}
    \end{subfigure}
    \begin{subfigure}{0.16\linewidth}
        \includegraphics[width=\textwidth]{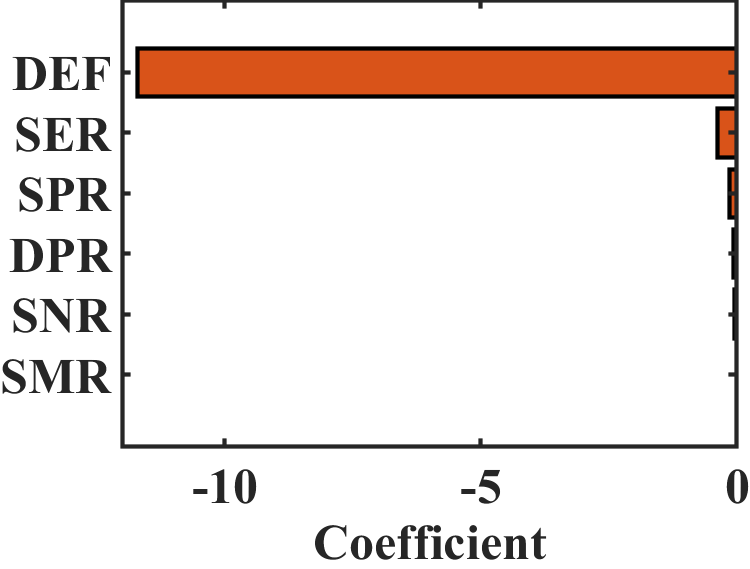}
        \caption{Model-0011}
        \label{xai3}
    \end{subfigure}
    \begin{subfigure}{0.16\linewidth}
        \includegraphics[width=\textwidth]{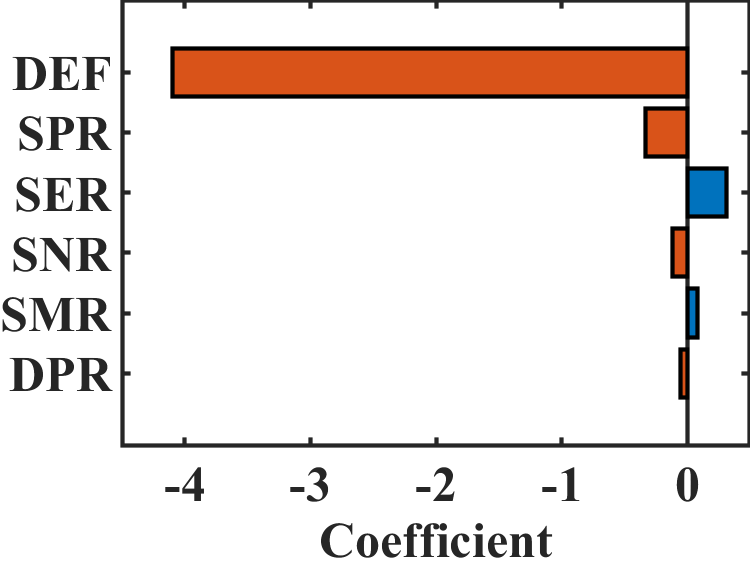}
        \caption{Model-0101}
        \label{xai4}
    \end{subfigure}
    \begin{subfigure}{0.16\linewidth}
        \includegraphics[width=\textwidth]{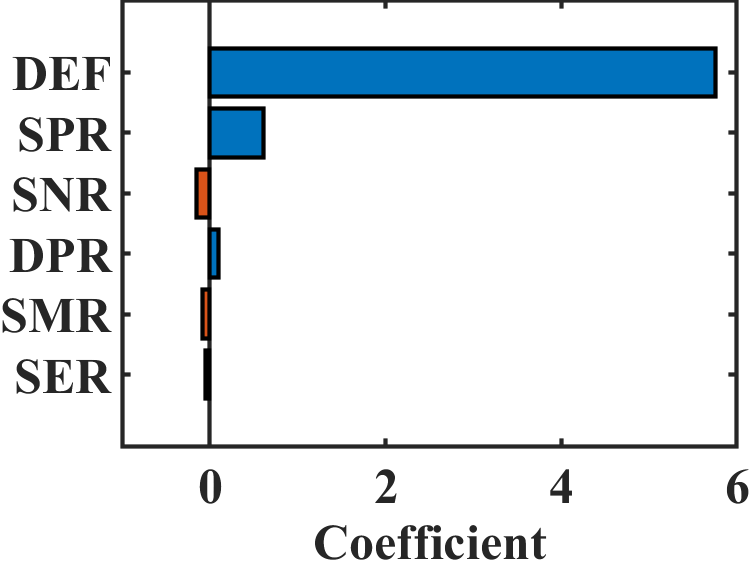}
        \caption{Model-0110}
        \label{xai5}
    \end{subfigure}
    \begin{subfigure}{0.16\linewidth}
        \includegraphics[width=\textwidth]{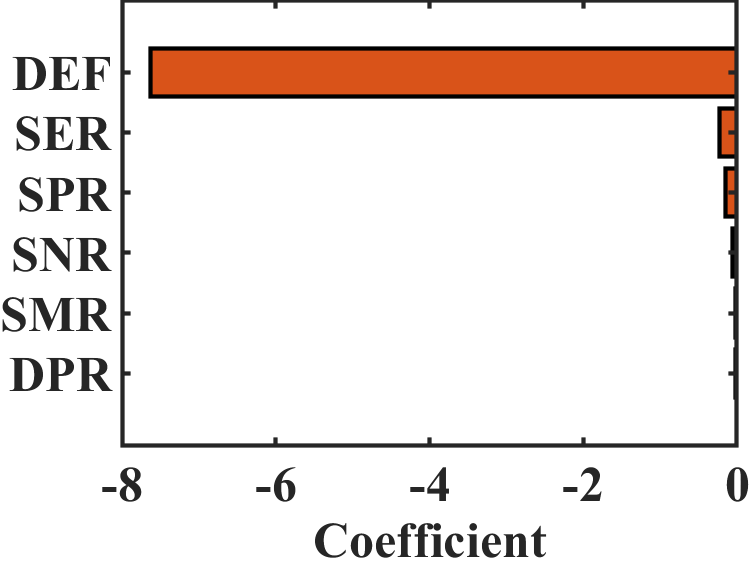}
        \caption{Model-0111}
        \label{xai6}
    \end{subfigure}
    \begin{subfigure}{0.16\linewidth}
        \includegraphics[width=\textwidth]{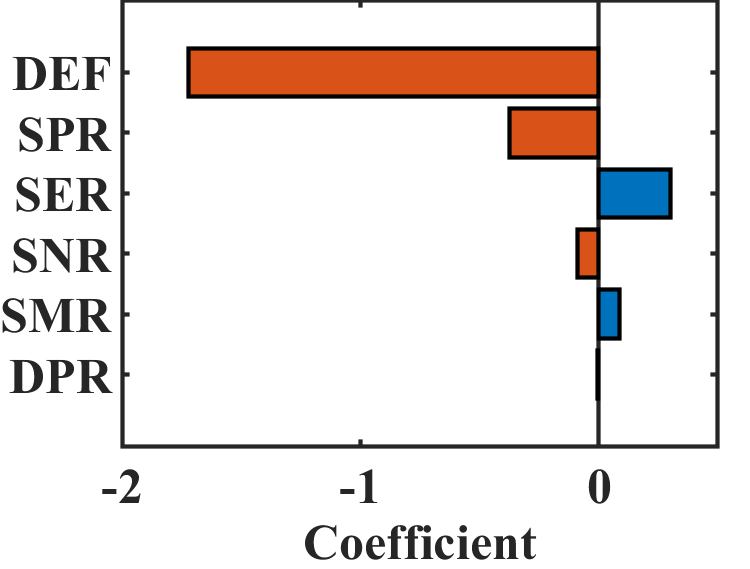}
        \caption{Model-1001}
        \label{xai7}
    \end{subfigure}
    \begin{subfigure}{0.16\linewidth}
        \includegraphics[width=\textwidth]{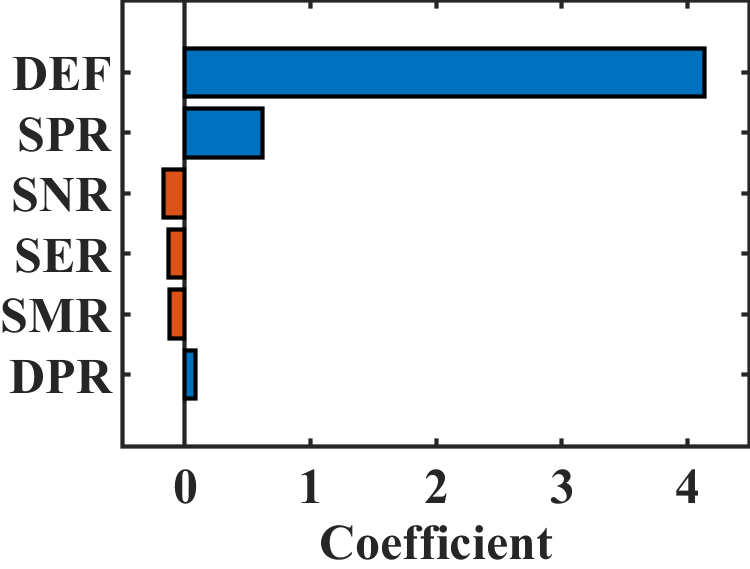}
        \caption{Model-1010}
        \label{xai8}
    \end{subfigure}
    \begin{subfigure}{0.16\linewidth}
        \includegraphics[width=\textwidth]{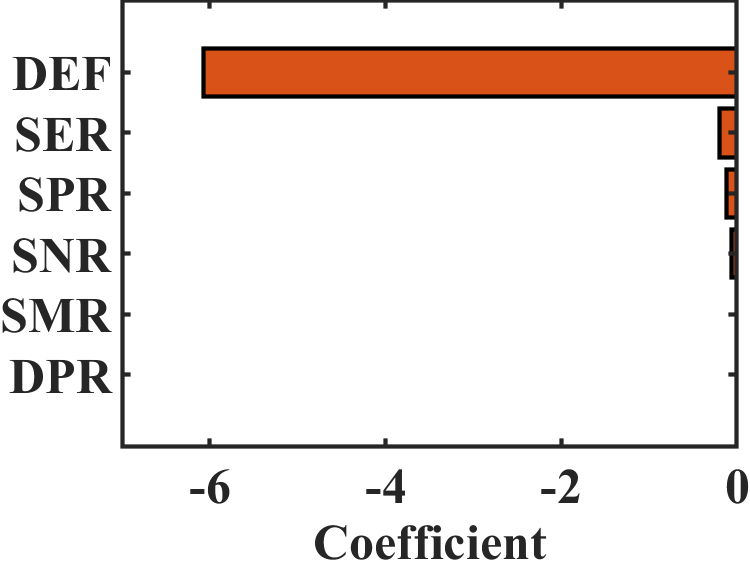}
        \caption{Model-1011}
        \label{xai9}
    \end{subfigure}
    \begin{subfigure}{0.16\linewidth}
        \includegraphics[width=\textwidth]{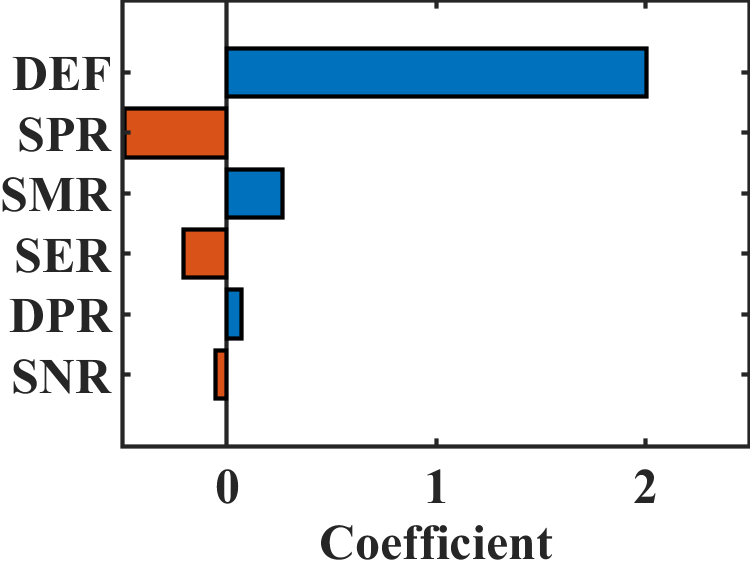}
        \caption{Model-1101}
        \label{xai10}
    \end{subfigure}
    \begin{subfigure}{0.16\linewidth}
        \includegraphics[width=\textwidth]{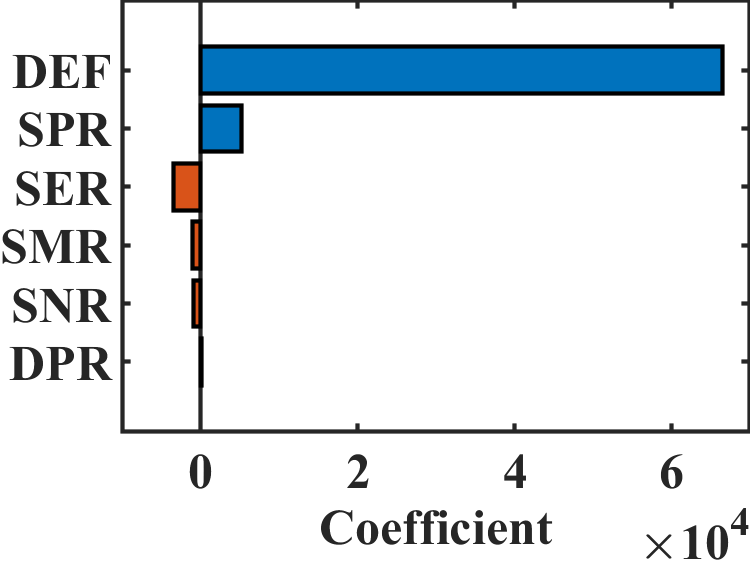}
        \caption{Model-1110}
        \label{xai11}
    \end{subfigure}
    \begin{subfigure}{0.16\linewidth}
        \includegraphics[width=\textwidth]{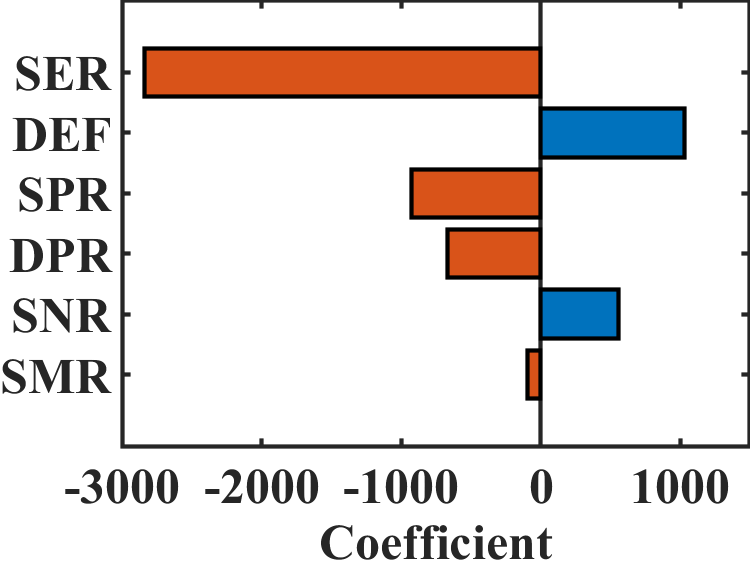}
        \caption{Model-1111}
        \label{xai12}
    \end{subfigure}
    \caption{Outputs of LIME for various models}
    \label{xai}
\end{figure}
The reward function is designed to be goal-oriented. Since the goal is to clean the signal effectively, the reward of +2 is awarded when the states corresponding to the filtered signal are identified as clean by the MLA, else a reward of 0 is given when the MLA identifies the states as noisy. To improve the MLA's speed and accuracy of identification, a LIME technique of explainable artificial intelligence (XAI) is integrated. The outputs from LIME (refer, to Fig.~\ref{xai}) suggest that in some cases few out of the six states/features can be used for noise identification and improve the accuracy of that model~\cite{lime, lime1, lime2}. The length of the bar in the figure represents that feature's ability to improve the identification of either noise (red bar) or clean (blue bar). From the figure, it can be safely assumed that the deformation ratio (DEF) is a major influence in identifying noise from clean signals at almost all levels.
Every action affects the environment differently, therefore it is necessary to find a fitting MLA for each action in every environment. Therefore, different MLAs are required to identify noise after a particular action is taken at a particular noise level. Six of the widely used MLAs are considered to achieve noise identification. They include support vector machine (SVM)~\cite{svm}, linear discriminant analysis (LDA)~\cite{lda}, artificial neural network (NN)~\cite{nn}, decision tree, logistic regression, k-nearest neighbor. They are trained and tested on the known training dataset before and after applying filters. The output label is either clean or noisy. In the training dataset, the label is clean if the actions corresponding to MOA, PLI, and WGN noise are HPF, NF, and LPF, respectively. All the other combinations of actions and noise are labeled as noisy. The MLA with the highest accuracy is selected and integrated with LIME to determine the most influencing features to improve accuracy. As a result, the final selected MLAs and corresponding features input for reward generation in each category are shown in Table~\ref{tab1}. The complete rewarding strategy is explained in Fig.~\ref{rwds}. Here, the affected state refers to the states of the environment at the current time step after a particular filtering action is executed. The MLA select line is a four-bit binary select line that uses only 12 states to select MLA for a particular action at a particular noise level.
\begin{table}[h]
    \centering
    \caption{MLA employed for reward generation due to each action at various noise levels.}
    \label{tab1}
    \begin{tabular}{c|c|c|c|c|c}
    \hline
       \textbf{$S_1S_2$} & \textbf{Noise Level} & \textbf{Action} & \textbf{Extracted Features} & \textbf{MLA} & \textbf{Model Accuracy} \\
       \textbf{$S_3S_4$} & \textbf{(in dB)} & & & & \textbf{(in \%)} \\
       \hline \hline
        0001 & -5 & HPF & All & NN  & 89.1 \\
        0010 & -5 & NF  & DEF, SER, SMR, SPR & SVM & 90.2 \\
        0011 & -5 & LPF & DEF, DPR, SER, SPR & SVM & 76.4 \\
        0101 & -1 & HPF & All & NN  & 96.6 \\
        0110 & -1 & NF  & All & NN  & 98.9 \\
        0111 & -1 & LPF & All & SVM & 94.8 \\
        1001 & +1 & HPF & All & NN  & 97.7 \\
        1010 & +1 & NF  & All & SVM & 98.9 \\
        1011 & +1 & LPF & All & SVM & 96   \\
        1101 & +5 & HPF & All & SVM & 73   \\
        1110 & +5 & NF  & All & SVM & 72.2 \\
        1111 & +5 & LPF & All & LDA & 64.9 \\ \hline
    \end{tabular}
\end{table}
\subsection{Supervised Reward based Deep Q-Network (supDQN) Agent} \label{sec:dqn}
The Deep Q-learning (DQN) is the extension of the Q-learning algorithm. It combines deep learning with the existing knowledge of Q-table update~\cite{dqn}. When the state space is continuous, it becomes difficult to contend all the states in a Q-table. In such cases, a universal function approximator like a deep neural network can be considered to estimate the Q-values. This function approximator is generally parameterized in $\theta$ and it learns to approximate much better by minimizing a loss as given in Eq.~\eqref{e4}
\begin{subequations} \label{e4}
\begin{align}
    L_i(\theta_i) & = \mathbb{E}_{s,a,r,s'\sim p(.)}((y_i - Q(s,a;\theta_i))^2) \\
    y_i & = r + \gamma \max_{a'}Q(s',a'; \theta_{i-1})
\end{align}
\end{subequations}
Where $L_i(\theta_i)$ is the loss function for the Q-value approximator network. It is equal to the expected temporal difference of the estimated Q-value in two successive iterations ($i$). $y_i$ represents the Q-value estimation parameterized over $\theta_{i-1}$ ($\theta$ at previous iteration) and $\gamma$ is the discount factor. $Q(s,a;\theta)$ represents the Q-value function of state $s$ and action $a$, parameterized in $\theta$.

\begin{figure}[h]
    \centering
    \includegraphics[width=0.8\textwidth]{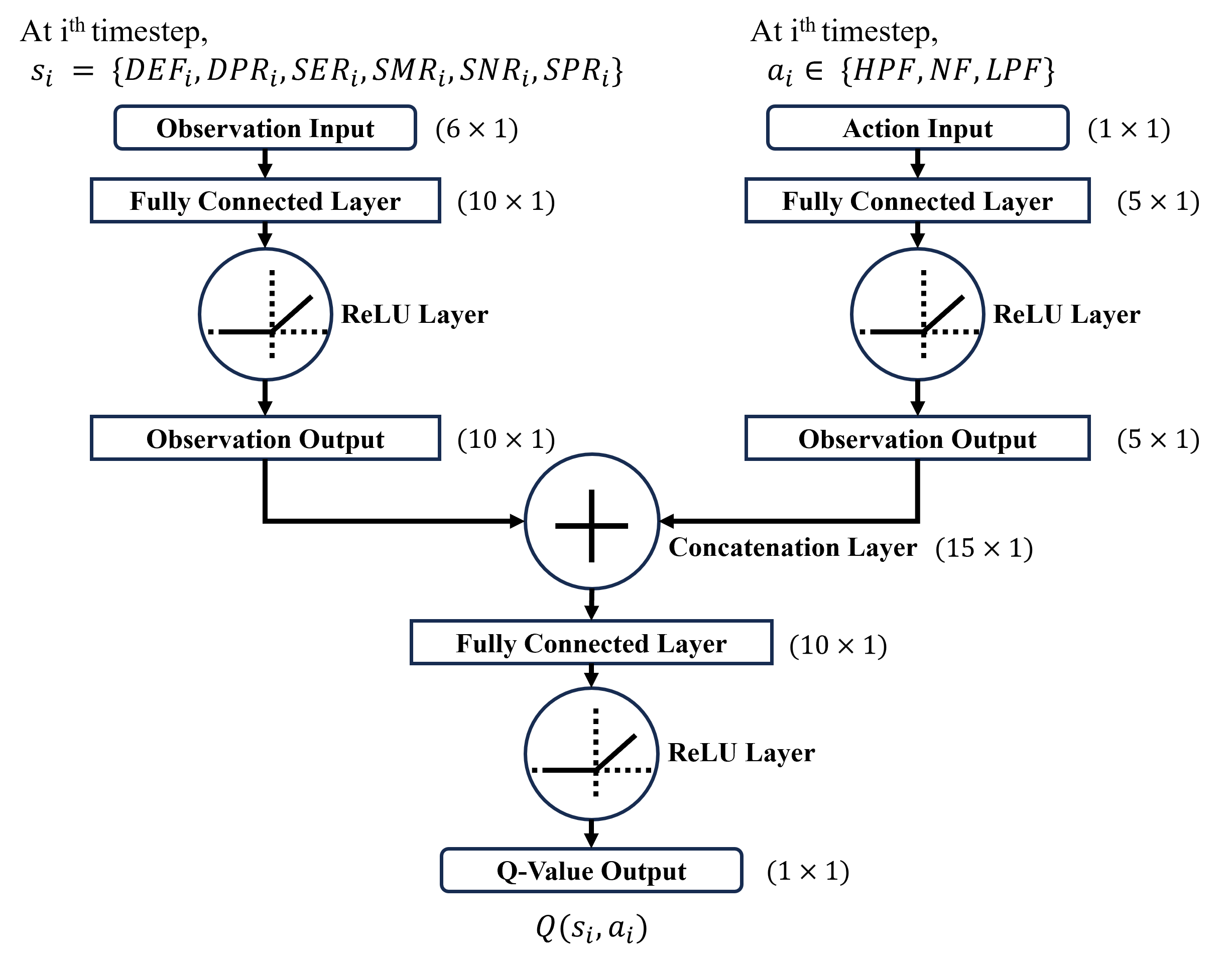}
    \caption{Neural Network for Q-value Estimation}
    \label{fig:critic}
\end{figure}

In this study, while the states for the training environment (ND1) are fixed and discrete in 64 timesteps, a Q-learning algorithm is suitable for learning the optimal policy to reduce noise. However, as the environment changes during testing or implementation, the Q-table fails to adjust the new states in the Q-table because the states are continuous in this case. Which results in the failure of the agent. In such cases, the Q-value is estimated using a deep neural network for a continuous state/observation (s) and discrete action (a).
\begin{equation} 
    Q^*(s,a;\theta) = \mathbb{E}[r+\gamma\max_{a'}Q^*(s',a';\theta)]
\end{equation}
$Q^*(s,a;\theta)$ is the quality of taking action $a$, starting from state/observation $s$ and following the optimal policy path thereafter parameterized over $\theta$ (also known as optimal Q-value function). Mathematically this is equivalent to the expected immediate return ($r$) plus the discounted future optimal return ($\gamma\max_{a'}Q^*(s',a';\theta)$). Here, $s'$ represents states/observations at the next timestep which is different from affected states explained in Section~\ref{sec:rwd}. Since the DQN agent is driven by a supervised learning-based rewarding strategy, it is named supDQN. Once the supDQN is trained, the Q-value estimator network stops updating its parameter $\theta$. When this trained model is simulated, the trained estimator network (shown in Fig.~\ref{fig:critic}) directly estimates the Q-value from the input states and actions and follows the path of the learned policy. As it tries to maximize the return, the signal is more filtered.

\subsection{Agent Performance Evaluation}
In Section~\ref{sec:noiseAddition}, noises are introduced in specified segments. The sequence of action being taken at each segment is noted and checked to see if the specific action is helpful in reducing the noise. In Section~\ref{sec:action}, intuitively it is found that action HPF, LPF, and NF are best equipped against MOA, WGN, and PLI respectively. Therefore, during the study, a desired set of actions is already known against the type of noise added at every segment. While evaluating the trained supDQN agent it is checked whether the agent acts according to the desired set of actions. The accuracy of the agent is calculated using the following equation
\begin{equation}
    Acc = \frac{Total\:\:timesteps-number\:of\:missed\:actions}{Total\:\:timesteps}\times100 
\end{equation}
Here, missed actions are the actions that deviate from the desired actions at any timestep. The missed actions and the agent's precision can be judged by using confusion matrices. The confusion matrix shows how the agent takes action against a particular type of noise in an environment. Noise 1, 2, and 3 correspond to MOA, PLI, and WGN respectively whereas actions 1, 2, and 3 correspond to HPF, NF, and LPF respectively. Similarly, an action plot is plotted against the timesteps in the x-axis. In the action plot, if the orange line overlaps completely with the blue line, then the action taken is the same as the desired action. In the filtered signal plot, the noisy signal after being filtered by the supDQN agent (orange) is compared against the original noisy signal (blue).

This filtering technique is compared against conventionally used filtering methods based on an error parameter ($\Omega$) which is calculated as follows
\begin{equation}\label{eq:rmse}
    \Omega = \frac{\sqrt{\frac{1}{N}\sum_{i=1}^N(y_i - x_i)^2}}{\sqrt{\frac{1}{N}\sum_{i=1}^N(n_i - x_i)^2}}
\end{equation}
Where, $y_i$, $x_i$, and $n_i$ represent the $i^{th}$ sample of the filtered signal, clean signal, and noisy signal respectively. A lower value of $\Omega$ indicates a more clean signal. The conventional filtering methods are compared against the proposed supDQN-filtering agent based on this parameter later in Table~\ref{tab:rmse}.
\section{Result \& Analysis} \label{resAnal}
In this section the output from the resulting supDQN filter is analyzed and compared with that of the conventional filters which include wavelet denoisers (with debauchies (dB4) and symlet (sym4) wavelet), low order high pass, low pass, and notch pass filter. To execute this the entire study is carried out in a 12\textsuperscript{th} generation Intel core i7 processor with an installed RAM of 16.0 GB capacity. MATLAB R2023a platform is used to execute all the programs and analyze results. Four cases for four different noise levels are prepared.

\subsection{Case I: Noise level -5 dB}
\begin{figure}[h]
\centering
\begin{subfigure}{0.4\linewidth}  
  \includegraphics[width=1\textwidth]{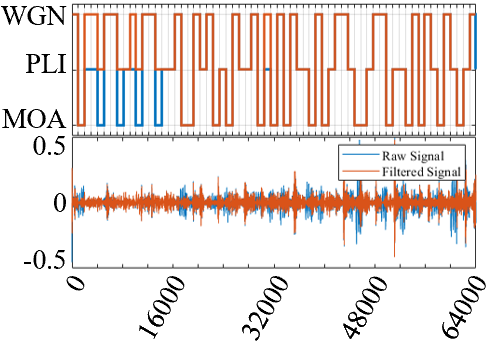}
  \caption{Action Plot}
  \label{n5db:1}
\end{subfigure}
\begin{subfigure}{0.4\linewidth}  
  \includegraphics[width=1\textwidth]{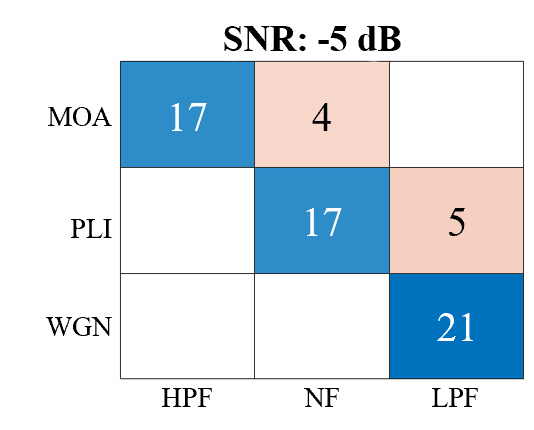}
  \caption{Confusion Matrix}
  \label{n5db:3}
\end{subfigure}
\begin{subfigure}{0.4\linewidth}  
  \includegraphics[width=1\textwidth]{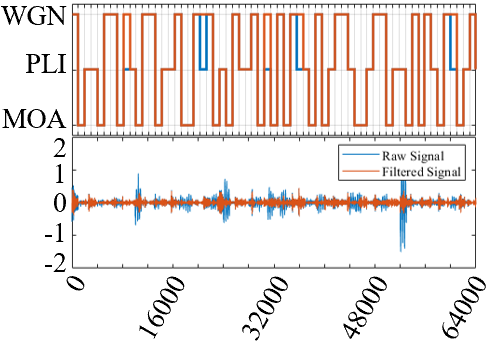}
  \caption{Action Plot}
  \label{n5db:4}
\end{subfigure}
\begin{subfigure}{0.4\linewidth}  
  \includegraphics[width=1\textwidth]{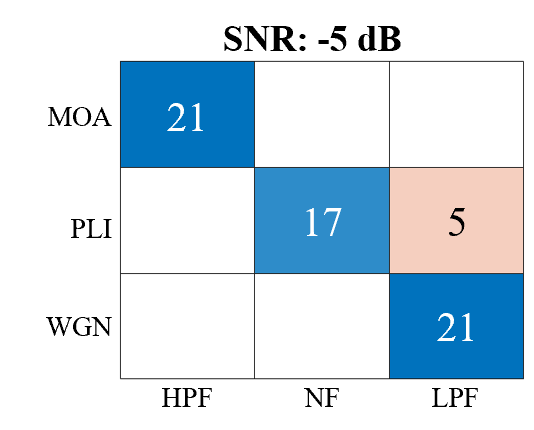}
  \caption{Confusion Matrix}
  \label{n5db:6}
\end{subfigure}
 \caption{Performance of supDQN agent at a noise level of -5 dB.} 
 \label{fig:n5db}
\end{figure}

In all the cases, the training environment is the ND1 dataset, and the rest of the environments (ND2 to ND9) are used as test environments for the validation of the agent. In this case, the noise is added such that the SNR corresponding to each segment is -5 dB. The three MLA for reward generation are referred from Table~\ref{tab1}. The agent-DQ1 is trained for 2000 episodes with each episode having a maximum of 64 timesteps. The learning rate of the deep neural network is 0.001 and an Adam optimizer with a gradient decay of 0.9 and $l^2norm$ gradient thresholding is used to optimize the weights of the neural network. The exploration rate is varied from 0.6 to 0.05 with a 0.003 decay at every timestep. The simulation on the testing dataset is observed and its performance analysis recorded while testing ND2 and ND3 is as shown in Fig.~\ref{fig:n5db}.
Fig.~\ref{n5db:1}-\ref{n5db:3} are the simulation results with the supDQN agent acting on dataset ND2 at a noise level of -5 dB. Whereas Fig.~\ref{n5db:4}-\ref{n5db:6} are the simulation results with supDQN agent acting on dataset ND3 at a noise level of -5 dB. Fig.~\ref{fig:n5db} suggests that at this noise level, the supDQN agent acts against the desired action 14 times while correctly identifying and filtering contaminant at 114 different timesteps. In the rest of the environments, only the accuracy and error ratio parameter $\Omega$ associated with the testing environment is mentioned in Table~\ref{tab:acc}. The optimizer settings for the estimator network and the exploration decay rate are the same for the rest of the cases. 

\subsection{Case II: Noise level -1 dB}
\begin{figure}[h]
 \centering
\begin{subfigure}{0.4\linewidth}  
  \includegraphics[width=1\textwidth]{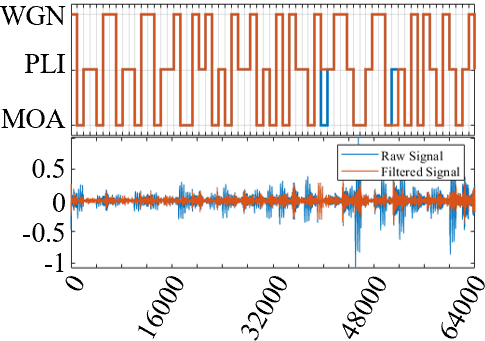}
  \caption{Action Plot}
  \label{n1db:1}
\end{subfigure}
\begin{subfigure}{0.4\linewidth}  
  \includegraphics[width=1\textwidth]{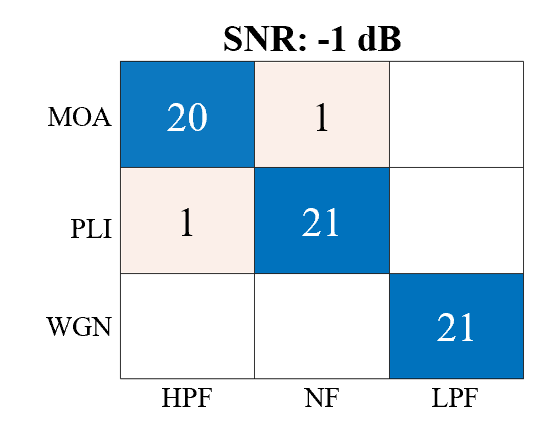}
  \caption{Confusion Matrix}
  \label{n1db:3}
\end{subfigure}
\begin{subfigure}{0.4\linewidth}  
  \includegraphics[width=1\textwidth]{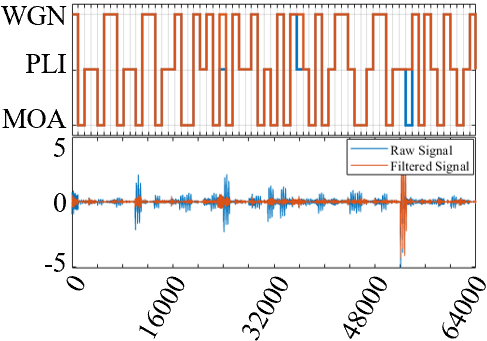}
  \caption{Action Plot}
  \label{n1db:4}
\end{subfigure}
\begin{subfigure}{0.4\linewidth}  
  \includegraphics[width=1\textwidth]{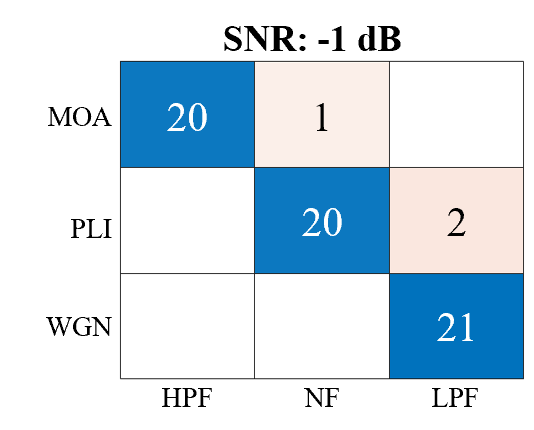}
  \caption{Confusion Matrix}
  \label{n1db:6}
\end{subfigure}
 \caption{Performance of supDQN agent at a noise level of -1 dB.} 
 \label{fig:n1db}
\end{figure}

The agent-DQ2 is trained and tested on the sEMG environment including a -1 dB level of noise at different segments. The simulation results are shown in Fig.~\ref{fig:n1db}. It suggests that the supDQN agent fails to follow the desired action in 5 cases and manages to correctly identify and decontaminate the signal in 123 different timesteps. 

\subsection{Case III: Noise level +1 dB}
\begin{figure}[h]
 \centering
\begin{subfigure}{0.4\linewidth}  
  \includegraphics[width=1\textwidth]{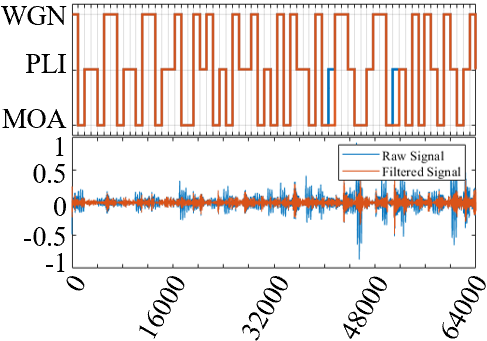}
  \caption{Action Plot}
  \label{p1db:1}
\end{subfigure}
\begin{subfigure}{0.4\linewidth}  
  \includegraphics[width=1\textwidth]{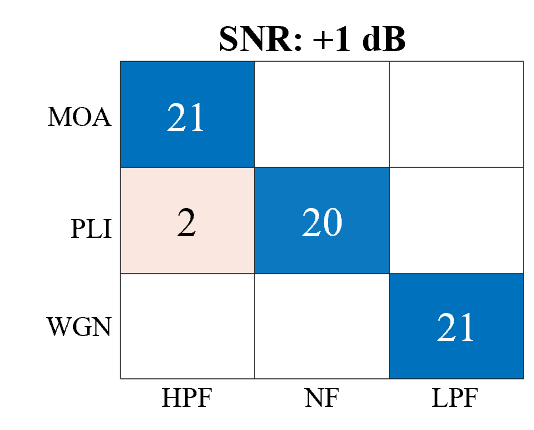}
  \caption{Confusion Matrix}
  \label{p1db:3}
\end{subfigure}
\begin{subfigure}{0.4\linewidth}  
  \includegraphics[width=1\textwidth]{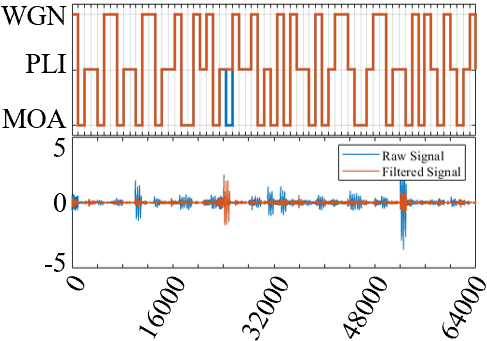}
  \caption{Action Plot}
  \label{p1db:4}
\end{subfigure}
\begin{subfigure}{0.4\linewidth}  
  \includegraphics[width=1\textwidth]{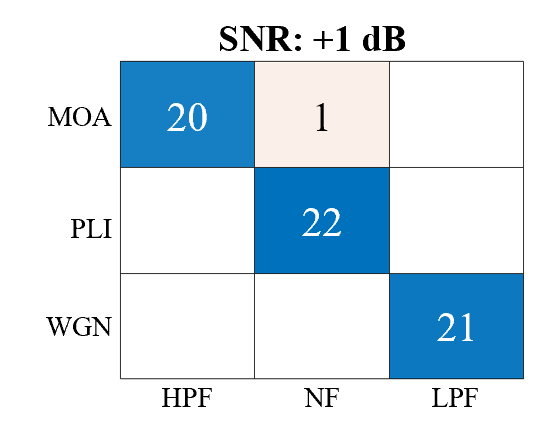}
  \caption{Confusion Matrix}
  \label{p1db:6}
\end{subfigure}
 \caption{Performance of supDQN agent at a noise level of +1 dB.} 
 \label{fig:p1db}
\end{figure}

The agent-DQ3 is trained and tested on an sEMG environment doped with a +1 dB level of noise at different segments. The simulation results are shown in Fig.~\ref{fig:p1db}. It suggests that the supDQN agent fails 4 times and correctly identifies and decontaminates the signal 124 times. 

\subsection{Case IV: Noise level +5 dB}
\begin{figure}[h]
 \centering
\begin{subfigure}{0.4\linewidth}  
  \includegraphics[width=1\textwidth]{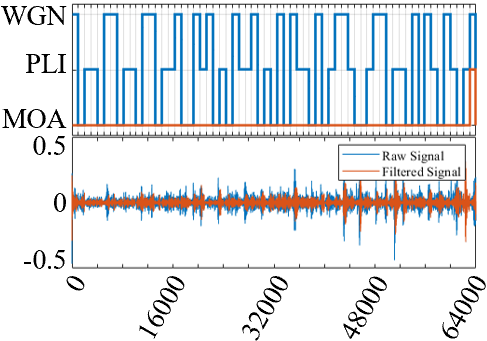}
  \caption{Action Plot}
  \label{p5db:1}
\end{subfigure}
\begin{subfigure}{0.4\linewidth}  
  \includegraphics[width=1\textwidth]{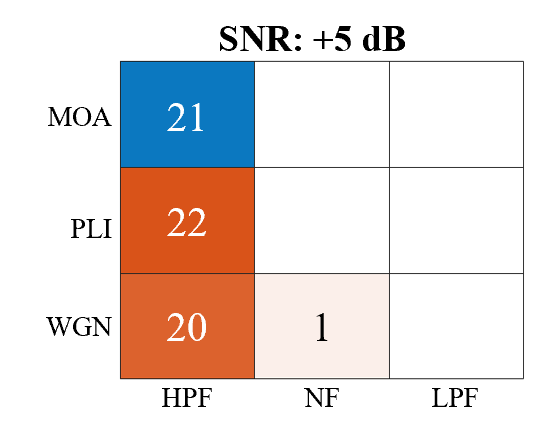}
  \caption{Confusion Matrix}
  \label{p5db:3}
\end{subfigure}
\begin{subfigure}{0.4\linewidth}  
  \includegraphics[width=1\textwidth]{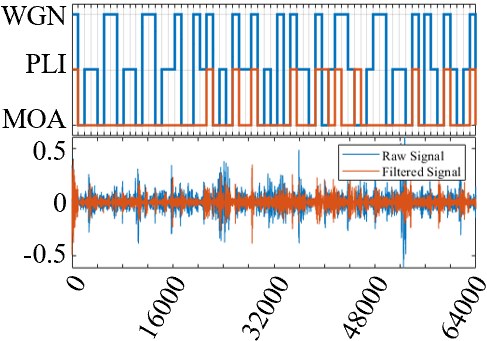}
  \caption{Action Plot}
  \label{p5db:4}
\end{subfigure}
\begin{subfigure}{0.4\linewidth}  
  \includegraphics[width=1\textwidth]{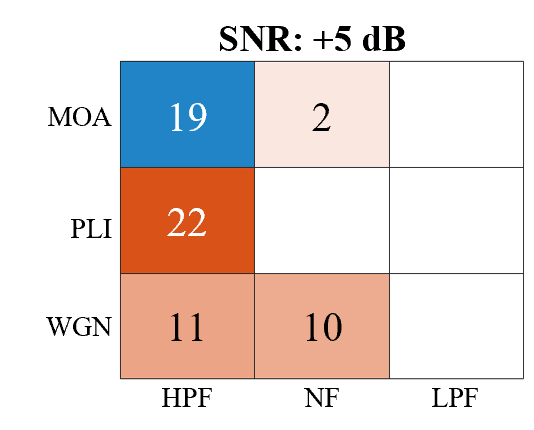}
  \caption{Confusion Matrix}
  \label{p5db:6}
\end{subfigure}
 \caption{Performance of supDQN agent at a noise level of +5 dB.} 
 \label{fig:p5db}
\end{figure}

The agent-DQ4 is trained and tested on an sEMG environment having a +5 dB level of noise at different segments. The simulation results are illustrated in Fig.~\ref{fig:p5db}. It suggests that the supDQN agent fails in 88 timesteps and succeeds in 40 timesteps. This happens due to the low impact of noise at this SNR level. At this level of SNR, it becomes difficult for machine learning agents to distinguish between noisy and clean signals based on the six features (DEF, DPR, SER, SMR, SNR, and SPR). For better distinction, there is a need for the exploitation of additional features that can prove to be helpful for noise identification at this noise level.

Table~\ref{tab:acc} presents the accuracy of supDQN agents in following the desired policy at different noise levels for all the testing datasets. 

\begin{table}[h]
    \centering
    \caption{Accuracy (in \%) of supDQN at following the desired actions in different noisy sEMG environments and at different noise levels}
    \label{tab:acc}
    \begin{tabular}{c|c|c|c|c|c|c|c|c}
    \hline
          & \textbf{ND2} & \textbf{ND3} & \textbf{ND4} & \textbf{ND5} & \textbf{ND6} & \textbf{ND7} & \textbf{ND8} & \textbf{ND9} \\ \hline \hline
         \textbf{-5dB} & 85.94 & 92.19 & 92.19 & 82.81 & 93.75 & 93.75 & 82.81 & 92.19 \\ 
         \textbf{-1dB} & 95.31 & 96.68 & 96.87 & 96.87 & 96.87 & 96.87 & 93.75 & 95.31 \\ 
         \textbf{+1dB} & 98.44 & 95.31 & 96.87 & 98.43 & 93.75 & 98.44 & 98.44 & 96.87 \\ 
         \textbf{+5dB} & 32.81 & 29.69 & 29.69 & 32.81 & 31.25 & 31.25 & 31.25 & 29.69 \\ \hline
    \end{tabular}
\end{table}
Table~\ref{tab:rmse} shows the performance of supDQN compared to the performance of other filters. The performance of each filter is evaluated using Eq.~\eqref{eq:rmse} at different noise levels for each test environment. The resulting values are then compared. Due to the consistency in maintaining the minimum $\Omega$ of 1.1974, which is expressed by the mean $\Omega$ in Table~\ref{tab:rmse}, it can safely be said that the supDQN decontaminates a mixed noise signal more effectively compared to the other filtering techniques. In 23 out of 32 cases, it performs better than the rest, and in 4 cases, it performs better than most filtering methods except HPF. In Fig.~\ref{fig:p5db}, it can be observed that when the agent is unable to identify noises correctly, it takes the HPF action more likely compared to the other. It is an adaptable learning agent that quickly adjusts to new environments after a little training since the learning process is dynamic.

\begin{table}[htbp]
\caption{Performance comparison table between the proposed method and conventional method}
\label{tab:rmse}
\centering
\begin{tabular}{c|c|c|c|c|c|c|c}
\hline
SNR & DS & \textbf{WD} & \textbf{WD} & \textbf{NF} & \textbf{LPF} & \textbf{HPF} & \textbf{supDQN} \\
 (dB)& &(db4)&(sym4) & & & & \\ \hline \hline
\multirow{8}{*}{\textbf{-5}} & \textbf{ND2} & 2.5925 & 2.5923 & 2.5896 & 2.5715 & 1.7632 & 1.0240 \\
 & \textbf{ND3} & 0.8647 & 0.8646 & 0.8637 & 0.8525 & 0.6875 & 0.5120 \\
 & \textbf{ND4} & 0.9995 & 0.9993 & 0.9962 & 1.2414 & 1.3021 & 0.8387 \\
 & \textbf{ND5} & 1.6508 & 1.6506 & 1.6483 & 1.6345 & 1.1602 & 0.7426 \\
 & \textbf{ND6} & 1.2050 & 1.2049 & 1.2034 & 1.1882 & 0.9835 & 0.6945 \\
 & \textbf{ND7} & 0.9996 & 0.9995 & 0.9984 & 1.1959 & 1.1369 & 0.7590 \\
 & \textbf{ND8} & 1.4331 & 1.4330 & 1.4315 & 1.4171 & 0.9654 & 0.6770 \\
 & \textbf{ND9} & 0.9659 & 0.9659 & 0.9648 & 0.9489 & 0.7590 & 0.5627 \\ \hline
\multirow{8}{*}{\textbf{-1}} & \textbf{ND2} & 0.7798 & 0.7797 & 0.7789 & 0.7735 & 0.5304 & 0.3954 \\
 
 & \textbf{ND3} & 0.2601 & 0.2601 & 0.2598 & 0.2564 & 0.2068 & 0.7543 \\
 & \textbf{ND4} & 0.9998 & 0.9997 & 0.9950 & 1.0016 & 0.4687 & 0.4334 \\
 & \textbf{ND5} & 0.4965 & 0.4965 & 0.4958 & 0.4916 & 0.3490 & 0.8684 \\
 & \textbf{ND6} & 0.3624 & 0.3624 & 0.3620 & 0.3574 & 0.2958 & 0.3209 \\
 & \textbf{ND7} & 0.9998 & 0.9998 & 0.9971 & 1.0002 & 0.4168 & 0.4642 \\
 & \textbf{ND8} & 0.4311 & 0.4310 & 0.4306 & 0.4263 & 0.2904 & 0.7407 \\
 & \textbf{ND9} & 0.2905 & 0.2905 & 0.2902 & 0.2854 & 0.2283 & 0.8037 \\ \hline
\multirow{8}{*}{\textbf{1}} & \textbf{ND2} & 0.9817 & 0.9816 & 0.9806 & 0.9738 & 0.6677 & 0.4523 \\
 & \textbf{ND3} & 0.3274 & 0.3274 & 0.3270 & 0.3228 & 0.2604 & 0.2877 \\
 & \textbf{ND4} & 0.9997 & 0.9996 & 0.9950 & 1.0203 & 0.5669 & 0.3772 \\
 & \textbf{ND5} & 0.6251 & 0.6250 & 0.6242 & 0.6189 & 0.4393 & 0.3377 \\
 & \textbf{ND6} & 0.4563 & 0.4563 & 0.4557 & 0.4499 & 0.3724 & 0.3210 \\
 & \textbf{ND7} & 0.9998 & 0.9997 & 0.9968 & 1.0064 & 0.5148 & 0.3652 \\
 & \textbf{ND8} & 0.5427 & 0.5427 & 0.5421 & 0.5366 & 0.3656 & 0.3215 \\
 & \textbf{ND9} & 0.3658 & 0.3657 & 0.3653 & 0.3593 & 0.2874 & 0.2959 \\ \hline
\multirow{8}{*}{\textbf{5}} & \textbf{ND2} & 8.1983 & 8.1976 & 8.1891 & 8.1319 & 5.5758 & 5.0203 \\
 
 & \textbf{ND3} & 2.7344 & 2.7343 & 2.7311 & 2.6957 & 2.1742 & 1.9753 \\
 & \textbf{ND4} & 0.9988 & 0.9984 & 1.0071 & 2.6387 & 4.0672 & 3.8499 \\
 & \textbf{ND5} & 5.2202 & 5.2198 & 5.2125 & 5.1687 & 3.6689 & 3.2906 \\
 & \textbf{ND6} & 3.8105 & 3.8103 & 3.8055 & 3.7575 & 3.1102 & 2.8730 \\
 & \textbf{ND7} & 0.9992 & 0.9989 & 1.0048 & 2.3431 & 3.5731 & 3.2845 \\
 & \textbf{ND8} & 4.5320 & 4.5317 & 4.5269 & 4.4814 & 3.0528 & 2.6743 \\
 & \textbf{ND9} & 3.0545 & 3.0543 & 3.0509 & 3.0007 & 2.4001 & 1.9990 \\ \hline

 \multicolumn{2}{c|}{\textbf{Mean $\Omega$}} & \textbf{1.5680} & \textbf{1.5679} & \textbf{1.5662} & \textbf{1.6609} & \textbf{1.3325} & \textbf{1.1974} \\ \hline

 \multicolumn{8}{c}{\textbf{Abbreviations:} DS: Dataset; WD: Wavelet denoisers} \\ \hline
\end{tabular}
\end{table}
\section{Conclusion} \label{conc}
This study presented a dynamic noise reduction technique that showed supremacy over other conventional filtering techniques. supDQN is an environment-interacting agent and its ability to filter segment by segment enables it to react to noise dynamically. It enables the filtering agent to filter noises dynamically. To achieve this, a supervised MLA was used to generate a reward and properly direct the supDQN agent to filter optimally in an unknown sEMG signal environment. At a higher level of noise in the signal (i.e. when SNR is -5 dB, -1 dB, or +1 dB) the MLA accurately identified clean from noisy signal which ensured a better degree of contamination removal. However, when the noise level in the signal is low (i.e. when SNR is +5 dB or more) the noisy signal is similar to the clean signal which causes difficulty for the MLA to distinguish noisy from the clean signal, and therefore, the agent fails to follow the desired filtering actions at every timestep. However, it still managed to minimize noise better in comparison to conventional techniques (refer Table~\ref{tab:rmse}). The proposed filtering technique in this study still lacks the confidence to achieve filtering in minimal time. However, efforts have been made in this study, to reduce the time consumption of the process by replacing higher-order filters with lower-order filters. In the future, this work can be modified to achieve faster filtering such that it becomes feasible to be used in prostheses, additionally, improvement in the rewarding strategy could be done which can identify the contribution of different noises in a noise-overlapped environment. The outcomes of this study can however be used in applications with the requirement of a clean sEMG signal without time constraints. Such as extended analytical research on clean sEMG signal, rehabilitation, clinical study, medical condition diagnosis, etc. In the future, this study can be extended to achieve faster filtering or filtering at lower levels of noise. Such that it can be integrated with prosthetic devices.

\section*{Ethics Approval}
This study was performed in accordance with the Declaration of Helsinki. This study on humans was approved by Malaviya National Institute of Technology Jaipur. All the adult participants provided their written informed consent to participate in this study.

\bibliographystyle{ieeetr}
\bibliography{ref}

\begin{thebibliography}{10}

\bibitem{WOS}
C.~PLC, ``Statistical data of semg related research: Web of science,'' 1997.

\bibitem{appl1}
D.~F. Stegeman, J.~H. Blok, H.~J. Hermens, and K.~Roeleveld, ``Surface emg models: properties and applications,'' {\em Journal of Electromyography and Kinesiology}, vol.~10, no.~5, pp.~313--326, 2000.

\bibitem{appl2}
G.~Boccia, D.~Dardanello, V.~Rosso, L.~Pizzigalli, and A.~Rainoldi, ``The application of semg in aging: a mini review,'' {\em Gerontology}, vol.~61, no.~5, pp.~477--484, 2015.

\bibitem{appl3}
C.~Frigo and P.~Crenna, ``Multichannel semg in clinical gait analysis: a review and state-of-the-art,'' {\em Clinical Biomechanics}, vol.~24, no.~3, pp.~236--245, 2009.

\bibitem{rmoa}
M.~B.~I. Reaz, M.~S. Hussain, and F.~Mohd-Yasin, ``Techniques of emg signal analysis: detection, processing, classification and applications,'' {\em Biological procedures online}, vol.~8, pp.~11--35, 2006.

\bibitem{moaWgn}
C.~J. De~Luca, L.~D. Gilmore, M.~Kuznetsov, and S.~H. Roy, ``Filtering the surface emg signal: Movement artifact and baseline noise contamination,'' {\em Journal of biomechanics}, vol.~43, no.~8, pp.~1573--1579, 2010.

\bibitem{pli}
J.~M. DeFreitas, T.~W. Beck, and M.~S. Stock, ``Comparison of methods for removing electromagnetic noise from electromyographic signals,'' {\em Physiological measurement}, vol.~33, no.~2, p.~147, 2012.

\bibitem{crossTalk}
C.~J. De~Luca, M.~Kuznetsov, L.~D. Gilmore, and S.~H. Roy, ``Inter-electrode spacing of surface emg sensors: reduction of crosstalk contamination during voluntary contractions,'' {\em Journal of biomechanics}, vol.~45, no.~3, pp.~555--561, 2012.

\bibitem{43}
A.~Phinyomark, C.~Limsakul, and P.~Phukpattaranont, ``A comparative study of wavelet denoising for multifunction myoelectric control,'' in {\em 2009 international conference on computer and automation engineering}, pp.~21--25, IEEE, 2009.

\bibitem{47}
A.~O. Andrade, S.~Nasuto, P.~Kyberd, C.~M. Sweeney-Reed, and F.~Van~Kanijn, ``Emg signal filtering based on empirical mode decomposition,'' {\em Biomedical Signal Processing and Control}, vol.~1, no.~1, pp.~44--55, 2006.

\bibitem{50}
L.~Liu, P.~Liu, E.~A. Clancy, E.~Scheme, and K.~B. Englehart, ``Signal whitening preprocessing for improved classification accuracies in myoelectric control,'' in {\em 2011 IEEE 37th Annual Northeast Bioengineering Conference (NEBEC)}, pp.~1--2, IEEE, 2011.

\bibitem{45}
C.~Tapia, O.~Daud, and J.~Ruiz-del Solar, ``Emg signal filtering based on independent component analysis and empirical mode decomposition for estimation of motor activation patterns,'' {\em Journal of Medical and Biological Engineering}, vol.~37, pp.~140--155, 2017.

\bibitem{46}
R.~M. Howard, R.~Conway, and A.~J. Harrison, ``An exploration of eliminating cross-talk in surface electromyography using independent component analysis,'' in {\em 2015 26th Irish Signals and Systems Conference (ISSC)}, pp.~1--6, IEEE, 2015.

\bibitem{51}
P.~Parker, K.~Englehart, and B.~Hudgins, ``Myoelectric signal processing for control of powered limb prostheses,'' {\em Journal of electromyography and kinesiology}, vol.~16, no.~6, pp.~541--548, 2006.

\bibitem{commonFilt}
J.~Wu, X.~Li, W.~Liu, and Z.~J. Wang, ``semg signal processing methods: A review,'' in {\em Journal of Physics: Conference Series}, vol.~1237, p.~032008, IOP Publishing, 2019.

\bibitem{steps}
M.~C. Tosin and A.~Balbinot, ``Identification and removal of contaminants in semg recordings through a methodology based on fuzzy inference and actor-critic reinforcement learning,'' {\em Expert Systems with Applications}, vol.~206, p.~117772, 2022.

\bibitem{biopac}
``Data acquisition, loggers, amplifiers, transducers, electrodes: Biopac.''

\bibitem{as1}
A.~Vijayvargiya, V.~Gupta, R.~Kumar, N.~Dey, and J.~M. R.~S. Tavares, ``A hybrid wd-eemd semg feature extraction technique for lower limb activity recognition,'' {\em IEEE Sensors Journal}, vol.~21, no.~18, pp.~20431--20439, 2021.

\bibitem{cc}
C.~Sinderby, L.~Lindstrom, and A.~Grassino, ``Automatic assessment of electromyogram quality,'' {\em Journal of Applied Physiology}, vol.~79, no.~5, pp.~1803--1815, 1995.

\bibitem{rwgn}
M.~Boyer, L.~Bouyer, J.-S. Roy, and A.~Campeau-Lecours, ``Reducing noise, artifacts and interference in single-channel emg signals: A review,'' {\em Sensors}, vol.~23, no.~6, p.~2927, 2023.

\bibitem{lime}
M.~T. Ribeiro, S.~Singh, and C.~Guestrin, ``" why should i trust you?" explaining the predictions of any classifier,'' in {\em Proceedings of the 22nd ACM SIGKDD international conference on knowledge discovery and data mining}, pp.~1135--1144, 2016.

\bibitem{lime1}
A.~Vijayvargiya, P.~Singh, R.~Kumar, and N.~Dey, ``Hardware implementation for lower limb surface emg measurement and analysis using explainable ai for activity recognition,'' {\em IEEE Transactions on Instrumentation and Measurement}, vol.~71, pp.~1--9, 2022.

\bibitem{lime2}
A.~Vijayvargiya, A.~Raghav, A.~Bhardwaj, N.~Gehlot, and R.~Kumar, ``A lime-based explainable machine learning technique for the risk prediction of chronic kidney disease,'' in {\em 2023 International Conference on Computer, Electronics and Electrical Engineering and their Applications (IC2E3)}, pp.~1--6, 2023.

\bibitem{svm}
G.~D. Fraser, A.~D.~C. Chan, J.~R. Green, and D.~T. MacIsaac, ``Automated biosignal quality analysis for electromyography using a one-class support vector machine,'' {\em IEEE Transactions on Instrumentation and Measurement}, vol.~63, no.~12, pp.~2919--2930, 2014.

\bibitem{lda}
F.~Duan, X.~Ren, and Y.~Yang, ``A gesture recognition system based on time domain features and linear discriminant analysis,'' {\em IEEE Transactions on Cognitive and Developmental Systems}, vol.~13, no.~1, pp.~200--208, 2021.

\bibitem{nn}
G.~Zhang, ``Neural networks for classification: a survey,'' {\em IEEE Transactions on Systems, Man, and Cybernetics, Part C (Applications and Reviews)}, vol.~30, no.~4, pp.~451--462, 2000.

\bibitem{dqn}
V.~Mnih, K.~Kavukcuoglu, D.~Silver, A.~Graves, I.~Antonoglou, D.~Wierstra, and M.~Riedmiller, ``Playing atari with deep reinforcement learning,'' {\em arXiv preprint arXiv:1312.5602}, 2013.

\end{thebibliography}

\clearpage

\end{document}